\documentclass[12pt,epsf]{article}
\usepackage{graphicx, amsmath}

\begin{document}
\sloppy
\begin{flushright}{SIT-HEP/TM-58}
\end{flushright}
\vskip 1.5 truecm
\centerline{\large{\bf Entropy production and curvature perturbation 
from dissipative curvatons}}
\vskip .75 truecm
\centerline{\bf Tomohiro Matsuda\footnote{matsuda@sit.ac.jp}}
\vskip .4 truecm
\centerline {\it Laboratory of Physics, Saitama Institute of Technology,}
\centerline {\it Fusaiji, Okabe-machi, Saitama 369-0293, 
Japan}
\vskip 1. truecm

\makeatletter
\@addtoreset{equation}{section}
\def\theequation{\thesection.\arabic{equation}}
\makeatother
\vskip 1. truecm

\begin{abstract}
\hspace*{\parindent}
Considering the curvaton field that follows dissipative slow-roll
 equation, we show that the field can lead to entropy production and
 generation of curvature perturbation after reheating. 
Spectral index is calculated to discriminate warm and thermal scenarios
 of dissipative curvatons from the standard curvaton model.
In contrast to the original curvaton model, quadratic potential is not
needed in the dissipative scenario, since the growth in the oscillating
period is not essential for the model.

\end{abstract}

\newpage
\section{Introduction}
Inflation and reheating are
important cosmological events in the early Universe.
Most models of inflation consider a scalar field that rolls slowly 
during inflation.
The scalar field is called inflaton.
The vacuum energy associated with the inflaton potential causes
acceleration of the Universe expansion.
The Universe during inflation is supposed to be cold
because of the rapid red-shifting of the primeval radiation, except for 
the Hawking temperature that is intrinsic to de Sitter space.
Reheating is thus required for the cold inflationary scenario to recover
hot Universe after inflation. 
Since the reheating may cause thermal production of unwanted relics,
such as gravitinos or moduli fields in the supergravity and the
string theory, entropy production after reheating is sometimes very
useful in ensuring successful nucleosynthesis.
A significant example of such entropy production would be thermal
inflation \cite{thermal-original} in which the inflaton (this field
should be discriminated from 
the inflaton of the primordial inflation that causes the first reheating)
is trapped at the origin because of the symmetry restoration caused by
the thermal background created by the primordial inflaton.
Thermal inflation starts when the Universe cools down and eventually 
the vacuum energy of the thermal inflaton potential
dominates the energy density, and ends when the symmetry breaking begins.
During thermal inflation the temperature decreases rapidly due to the
accelerated expansion.
For the thermal inflation scenario 
small initial value of the field ($\phi<T_R$, where $T_R$ is
the reheating temperature) is
crucial for the thermal trapping, as the chaotic initial condition
 ($T_R\ll \phi \sim M_p $) may prevent the required symmetry restoration.
In this paper, we consider a complementary
scenario of thermal inflation in which the
field fails to cause symmetry restoration.
Considering the dissipative field motion after reheating, we show that
the scenario may also cause generation of curvature perturbation.
The situation is in contrast to the original thermal inflation scenario,
in which the generation of the curvature perturbation is not
significant.\footnote{This statement needs more explanations.
Cosmological phase transitions may occur not simultaneous in space, but
occur with time lags in different Hubble patches.
The time lags may arise from the long-wavelength inhomogeneities caused
by the light-field fluctuations that exit horizon during primordial
inflation. 
Then the inhomogeneous phase transition \cite{IH-pt} may cause generation
of the curvature perturbation at the end of thermal inflation.
In Ref. \cite{thermal-curvgen}, utilizing a fluctuating
coupling of a light field (flaton) with the fields in thermal bath,
it has been shown that a mechanism of generating the 
curvature perturbation at the end of thermal inflation may liberate
some inflation models and may cause significant non-gaussianity.}
The mechanism for generating curvature perturbation is similar to the
curvatons, although there are crucial differences that discriminate our
scenario from the original curvaton.
In this paper the subscript``1I'' is used for the first (primordial)
inflation stage and ``2I'' for the dissipative curvaton.

\section{Dissipative field in the Universe}
In this section we first review the dissipative field motion in the warm
inflation scenario. 
Then we consider the dissipative field motion of the curvaton field
$\phi_{2I}$ after reheating.  
It will be shown that the field $\phi_{2I}$ may lead to secondary
inflationary expansion
and generation of the curvature perturbation after reheating.

\subsection{Warm inflation (review)}

The inflation scenario with the dissipating inflaton field is known 
as warm inflation \cite{warm-inflation-original}, in which the
dissipation sources radiation and keeps the temperature
$T>H$ during inflation.
The condition $T>H$ characterizes the warm inflation scenario,
which includes weak dissipation (WD) scenario where the slow-roll is
{\bf not} due to the significant dissipation.
One of the important points that discriminate our scenario from the 
warm inflation scenario is the source of the radiation.   
In the warm inflation scenario the radiation energy density $\rho_R$ is
continuously sourced by the dissipation of the inflaton field.
The source mechanism leads to the situation that the time-derivative
$\dot{\rho}_R$ is 
negligible ($\dot{\rho}_R\sim 0$, or more precisely $\dot{\rho}_R \ll
4H\rho_R$) during inflation.\footnote{In contrast to the warm inflation
scenario, our scenario considers the radiation-dominated Universe from
the beginning, where the radiation is caused by the decay products of
the primordial inflaton field.
In the radiation-dominated Universe we have in mind the situation that
the dissipation from the curvaton field may source radiation, but its
amount may or may not be enough to sustain the radiation against
red-shifting.} 
A confusing aspect of the warm inflation scenario may be
that intuitively realizing significant radiation during inflation
seems very difficult.
On the other hand, if a field has time-dependent mass during inflation,
one intuitively knows that excitation of such field is inevitable.
The dissipation can be significant when the decay of the excitation
causes energy loss.
There may be some argument related to the thermalization of the decay
product, but the thermalization is not essential for the dissipative
curvaton model.
Further details are discussed in Appendix A, with regard to
thermalization and dissipation of the curvaton field.

Before discussing the slow-roll dynamics during the radiation-dominated
Universe, we first review the scenario of warm inflation, in which the
inflaton potential dominates the energy of the Universe and the
dissipation from the inflaton field sources the radiation.
Following the usual convention of warm inflation, the equation of motion
when the inflaton is dissipative is given by 
\begin{equation}
\label{eqofmo1st}
\ddot{\phi}_{1I}+3H(1+r)\dot{\phi}_{1I}+V(\phi_{1I},T)_{\phi_{1I}}=0,
\end{equation}
where the subscript denotes the derivative with respect to the field
$\phi_{1I}$.
Understanding the dissipative dynamics from the quantum
field theory is the challenge of realizing warm inflation.
Much work has been devoted to this problem in terms of analytic
approximations \cite{Gamma-Ms0, Gamma-Ms1, warm-analytic} 
and with numerical based methods \cite{numerical-based}.
The quasiparticle approximation is considered in
Ref. \cite{Gamma-Ms0,Gamma-Ms1}, which gives dissipative force in the
zero-temperature background.
And the equilibrium approximation, which
gives dissipation in the background with non-zero temperature, has been 
considered by many authors.
Obviously, the zero-temperature limit of the equilibrium approximation
does not lead to the quasiparticle approximation.
In these calculations
the strength of the damping (i.e., the frictional force) is measured by
the rate $r$ given by the dissipation coefficient $\Upsilon$ and the
Hubble parameter $H$;
\begin{equation}
r\equiv \frac{\Upsilon}{3H}.
\end{equation}
A short note on the dissipation coefficient $\Upsilon$
is added in the latter part of this section.
The effective potential of the field $\phi_{1I}$
 is expressed as $V(\phi_{1I},T)$,
which depends on the radiation temperature $T$. 
The dissipation coefficient $\Upsilon$ of the inflaton, which is related
to the microscopic physics of the interactions, leads to the energy
conservation equation;
\begin{equation}
\label{cons-rad}
\dot{\rho}_R+4H \rho_R = \Upsilon \dot{\phi}_{1I}^2,
\end{equation}
where $\rho_R$ is the energy density of the radiation and 
the right hand side ($ \Upsilon\dot{\phi}_{1I}^2$) represents
the source of the radiation from the dissipation.\footnote{The
slow-roll equation leads to the equation for the source term;  
$\Upsilon \dot{\phi}_{1I}^2 \simeq V_{\phi_{1I}}\dot{\phi}_{1I}
\simeq \dot{V}$.}
Considering the scale factor $a$ of the Universe, the radiation scales
as $\rho_R\propto 
a^{-4}$ when the source term is negligible, but $\rho_R$ may 
behave like a constant ($\dot{\rho}_R\simeq 0$) 
when the source term is significant ($4H\rho_R \sim \Upsilon
 \dot{\phi}_{1I}^2$)
in the energy conservation equation.
In this paper, the former will be denoted by the ``thermal phase'' 
and the latter will be denoted by the ``warm phase''.
The slow-roll equation in the thermal phase is the key topic in this
paper. 

Scenario with $r>1$ is usually called 
strongly dissipating (SD) scenario.
In this paper we consider only the SD motion with 
$r>1$, because we have in mind the situation where the conventional
slow-roll condition is already violated by the potential.
With regard to the dissipative field equation, the effective 
slow-roll parameters are different from the conventional
(non-dissipating) ones.  
They are given by
\begin{eqnarray}
\epsilon_w &\equiv& \frac{\epsilon}{(1+r)^2},\nonumber\\
\eta_w &\equiv& \frac{\eta}{(1+r)^2},
\end{eqnarray}
where the non-dissipating slow-roll parameters ($\epsilon$ and $\eta$)
are defined by 
\begin{eqnarray}
\epsilon&\equiv& \frac{M_p^2}{2}\left(\frac{V_\phi}{\rho_{Tot}}\right)^2,
\nonumber\\
\eta &\equiv& M_p^2\frac{V_{\phi\phi}}{\rho_{Tot}},
\end{eqnarray}
where $M_p\equiv (8\pi G)^{-1/2}\simeq 2.4\times 10^{18}$GeV is the
Planck mass and $\rho_{Tot}$ is the total energy density.
Here the subscript under $V$ denotes the derivative with respect to the
inflaton field $\phi$.
The warm-phase condition $4H\rho_R \simeq \Upsilon \dot{\phi}^2$
combined with the slow-roll equation leads to more
stringent conditions \cite{gil-berera}
\begin{eqnarray}
\label{slo-ro-warm}
\epsilon &<& (1+r)\nonumber\\
\eta &<& (1+r),
\end{eqnarray}
which is required for the slow-roll in the warm phase \cite{gil-berera}.
In the thermal phase, where the warm-phase condition is violated, we
introduce an alternative $\dot{V}< 4H V$, which suggests that the
potential decreases slower than the radiation.
The slow-roll conditions in the thermal phase cannot be the same as
in the warm phase. 

\begin{figure}[h]
 \begin{center}
\begin{picture}(200,140)(100,80)
\resizebox{15cm}{!}{\includegraphics{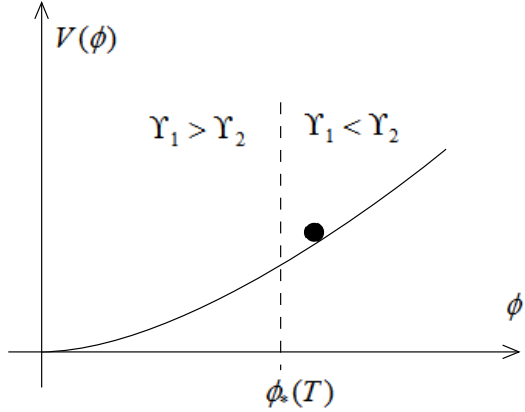}} 
\end{picture}
 \caption{$\Upsilon_1$ and $\Upsilon_2$ are calculated using the
  equilibrium and the quasiparticle approximations, respectively.
 The critical point $\phi_*$ is not a constant.
 It moves rapidly in the thermal phase because it depends crucially on
  the temperature.}    
\label{fig:coefficient}
 \end{center}
\end{figure}

To quantify the dissipation coefficient $\Upsilon$, 
it would be useful to consider past results for the typical
superpotential \cite{Gamma-Ms1, Moss:2006gt}.
In the warm inflationary scenario, final product of the dissipation is
excitation of light fermions, which may enhance the loop corrections to
the inflaton potential.
Then the interactions related to dissipation may lead to the
violation of the flatness conditions which inflation requires.
Because of this concern, one needs to consider the loop corrections when
the light fields thermalize during warm inflation.
The light sector, where the decay products thermalize during
inflation, typically have couplings representing self-interactions
or interactions with other light fields.
These interactions are implicitly assumed in addition to the basic
interactions which are typically given by
\begin{equation}
{\cal L}_I \simeq  g_1 \phi^2 \chi^2 -g_2 \chi^2 \psi^2,
\end{equation}
where the inflaton $\phi$ couples to the intermediate field $\chi$ and
finally decay into light fermions $\psi$.
Therefore, the relaxation time of the radiation may be independent of the
interactions related to the dissipation mechanism.
We thus ``assume'' thermalization in this paper.
For the adiabatic case, the dissipation coefficient usually depends on
the temperature $T$ and the expectation value of the field.
Considering interactions given by the superpotential
\begin{equation}
W=g_1\Phi X^2+g_2 XY^2,
\end{equation}
where $g_1$ and $g_2$ are coupling constants, and $\Phi, X, Y$ are
superfields whose scalar components are given by $\phi, \chi, y$.
Fermionic partners of the scalar components $\phi, \chi, y$ are 
$\psi_\phi, \psi_\chi, \psi_y$.
The supersymmetry is broken by the inflaton field, which leads to 
one-loop contribution to the inflaton potential.
Usually, inflation is associated with a very weakly coupled inflaton in
the inflation sector to protect the inflaton potential from large
quantum corrections. 
However, in some inflation models, for instance when hybrid-type
potential is considered, the interactions of the inflaton field may not
be negligible\cite{hybrid-radiative}.
In this case, supersymmetry can play significant role in inflationary
cosmology.
In fact, in supersymmetric models the restrictions on the couplings are
less severe due to the cancellations that may be still effective when
supersymmetry is softly\footnote{''Soft breaking'' does not mean that
the energy scale of the supersymmetry breaking is small compared with
the inflation scale.
Soft breaking occurs when ``soft terms'' break supersymmetry.
See textbook\cite{SUSY-text} for more details.} or thermally 
broken during inflation.
As the result, the gradient of the inflaton potential can be small
enough to allow 
inflation whilst the thermalization time for the radiation remains
short, allowing warm inflation for moderate values of the coupling
constants in these interactions.\footnote{See Ref.\cite{Moss:2006gt} for
more details.}
We consider the model in which the field motion ($\dot{\phi}\ne 0$) 
causes excitation of the heavy
intermediate field $\chi$ that has the mass proportional to $\phi$.
Then the heavy intermediate field decays into light fermions to
dissipate the energy from the field $\phi$. 
The situation (i.e., the heavy intermediate field $+$ light fermions
with $\dot{\phi}\ne 0$) is not peculiar for particlecosmology.
During warm inflation, when $\phi$ is large, the field $y$ and
its fermionic 
partner $\psi_y$ are massless in the global supersymmetric limit.
The mediating field is $\chi$, which obtains large mass 
$m_\chi\sim g_1 \phi$ from the interaction, and
the dissipation is caused by the excitation of the
mediating field $\chi$ that decays into massless fermions.
Considering the equilibrium approximation,
at high temperature ($m_\chi \ll T$) the dissipation coefficient is
given by $\Upsilon \propto (g_1^2/g_2^2)T$, while at low temperature
($m_x \gg T$) the coefficient is given by $\Upsilon \propto
g_2^4(T^3/\phi^2)$.
The dissipation coefficient must include terms both from 
the equilibrium approximation and the quasiparticle approximation 
at the same time.
However, in the conventional warm inflation scenario
the less effective one can be disregarded, since during warm inflation 
the changes in the inflaton field $\phi$ and the temperature $T$
are not significant due to the slow-roll conditions.
In contrast to the warm inflation scenario, the slow-roll equation for
the curvatons in the thermal phase should be evaluated with the 
significant change in the temperature $T$.
Therefore, for the dissipative curvaton scenario, 
the dissipation coefficient must include both terms from
the low-temperature and the quasiparticle approximations.
For instance, we consider $\Upsilon \simeq \Upsilon_1+\Upsilon_2 
\equiv C_1 T^3/\phi^2 + C_2 \phi$ for the dissipative motion 
after reheating,
where $\Upsilon_1$ is obtained using the  equilibrium approximation 
and is significant for $\phi < \phi_* \equiv T \times 
(C_1/C_2)^{1/3}$, while
 $\Upsilon_2$ is obtained using the quasiparticle approximation
and is significant for $\phi > \phi_*$.
See also Figure \ref{fig:coefficient}. 
Considering $N_\chi$ and $N_{\psi}$ for the number of intermediate
fields and the decay products, the coefficients $C_1$ 
and $C_2$ are typically given by
$C_1\sim 0.1 \times g_2^4 N_\chi N_\psi^2$ \cite{gil-berera}
and $C_2\sim 0.1 \times 
N_\chi/\sqrt{N_\psi}g_1^3/g_2$ \cite{Gamma-Ms0, Gamma-Ms1}.

\begin{figure}[h]
 \begin{center}
\begin{picture}(200,220)(100,360)
\resizebox{15cm}{!}{\includegraphics{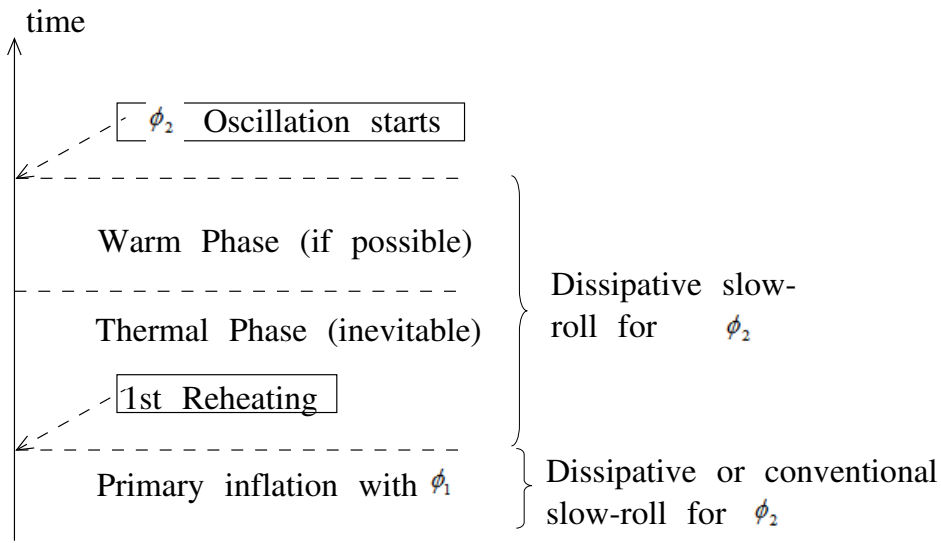}} 
\end{picture}
 \caption{Evolution of the Universe is shown for the dissipative
  scenario. The potential of the dissipative field $\phi_2$ may start
  dominating the Universe either in the thermal phase or in the warm
  phase. Then the second inflation (dissipative inflation) starts for
  the dissipative field $\phi_2$. Dissipative inflation in the warm
  phase may be called warm inflation, but the spectral index is
  different from the original scenario.
Therefore, we use ``dissipative inflation'' for the secondary inflation
stage even if it is in the warm phase.
 Thermal phase is inevitable for the modest scenario with $T_R>T_{eq}$.}  
\label{fig:1st}
 \end{center}
\end{figure}

A natural expectation from the standard warm inflation scenario
is that similar phenomenon (dissipative
slow-roll) should be observed in the evolution of other scalar fields
during and after inflation, either in the thermal or in the warm phases.
In this paper we apply this idea to the field motion of the
dissipative curvaton field $\phi_{2I}$ after reheating.
Evolution of the field before reheating depends crucially on the
inflation model, which may either be cold or warm.
Because of the crucial model-dependence of the scenario, we will not
discuss the evolution of the field before reheating.
Instead, we simply assume that the curvaton stays at $\phi_{2I}>T_R$
before reheating. 
See Figure.\ref{fig:1st} for the evolution of the Universe.

\subsection{Radiation and dissipation after reheating}

To understand the evolution of the dissipative field after
inflation, we consider $V(\phi_{2I})$, which is the monomial potential
given by 
\begin{equation}
V(\phi_{2I})=\lambda_n \frac{\phi_{2I}^n}{M_p^{n-4}},
\end{equation}
where the less effective terms are disregarded for simplicity.
The energy density of the Universe is given by
\begin{equation}
3 H^2 M_p^2\equiv \rho_{Tot} \simeq \rho_R + V(\phi_{2I}).
\end{equation}
The slow-roll field equation with $r\gg 1$ leads to
\begin{eqnarray}
\Upsilon \dot{\phi}_{2I}^2 &\simeq& \Upsilon
\left(\frac{V_{\phi_{2I}}}{3Hr}\right)^2
\nonumber\\
&\simeq& 
\left\{
\begin{array}{cccc}
\frac{(\phi_{2I} V_{\phi_{2I}})^2}{C_1 T^3} &=&\frac{n^2}{C_1}
\frac{V^2}{T^3}& \,\,\,\,\, (\phi_{2I}<\phi_*)\\
\frac{(\phi_{2I} V_{\phi_{2I}})^2}{C_2 \phi_{2I}^3}&
=&\frac{n^2}{C_2}
\frac{V^2}{\phi_{2I}^3} &\,\,\,\,\, (\phi_{2I}>\phi_*),
\end{array}
\right.
\end{eqnarray}
where $\phi_{2I} V_{\phi_{2I}}=nV(\phi_{2I})$ is used for the monomial
potential. 
If the radiation satisfies the warm-phase condition $\dot{\rho}_R
 < 4H\rho_R$, which leads to $\Upsilon \dot{\phi}_{2I}^2 \simeq
4H\rho_R$, the radiation of the Universe 
is sourced by the dissipation and the temperature in the warm phase 
is given by
\begin{eqnarray}
\label{sta-rad}
T^7&=&\alpha_1 \frac{V(\phi_{2I})^2}{H} \,\,\,\,\, (\phi_{2I}<\phi_*)\\
\label{sta-rad2}
T^4 &=& \alpha_2 \frac{V(\phi_{2I})^2}{H\phi_{2I}^3} 
 \,\,\,\,\, (\phi_{2I}>\phi_*),
\end{eqnarray}
where we introduced a new parameter $\alpha_{1,2}$ defined by
\begin{eqnarray}
\alpha_1 &\equiv& \frac{n^2}{4C_1 C_T} \sim 10^{-6}
\left(\frac{n}{4}\right)^2
\left(\frac{10^4}{C_1}\right)
\left(\frac{10^2}{C_T}\right)\\
\alpha_2 &\equiv& \frac{n^2}{4C_2 C_T} \sim 10^{-2}
\left(\frac{n}{4}\right)^2
\left(\frac{1}{C_2}\right)
\left(\frac{10^2}{C_T}\right).
\end{eqnarray}
Here $C_T$ is the constant that is defined as
$\rho_R \equiv C_T T^4$.
The warm phase must be an instantaneous phenomenon if it appears in the
radiation-dominated Universe, where the relation $\rho_R >V(\phi_{2I})$
is crucial.

\subsection{Slow-roll conditions for the dissipative field}
Considering the dissipative field equation, the effective 
slow-roll parameters are given by
\begin{eqnarray}
\label{slo-ro}
\epsilon_w &\equiv& \frac{1}{(1+r)^2}\frac{M_p^2}{2}
\left(\frac{V_\phi}{\rho_{Tot}}\right)^2<1\nonumber\\
\eta_w &\equiv& \frac{1}{(1+r)^2}M_p^2\frac{V_{\phi\phi}}{\rho_{Tot}}<1,
\end{eqnarray}
which are needed for the condition
$\ddot{\phi}_{2I}<(3H+\Upsilon)\dot{\phi}_{2I}$. 

In the warm phase, $\dot{\rho}_R < 4H\rho_R$ imposes more stringent
conditions \cite{gil-berera} $\epsilon < (1+r)$ and $\eta < (1+r)$.
In the thermal phase, the slow-roll conditions are $\epsilon_w<1$ and
$\eta_w<1$, which result in $\epsilon_w\le\eta_w<1$ for the monomial
potential. 
In addition to the conventional slow-roll conditions, there is an
obvious condition $\dot{V}<4HV$ in the thermal phase, which is needed to
ensure that the radiation decreases faster than the potential.
During the radiation-dominated epoch,  only the thermal phase is
relevant for the dynamics of $\phi_{2I}$
.

\subsection{After reheating ($\rho_{Tot}\simeq \rho_R$)}

After reheating the temperature decreases as $T\propto a^{-1}$.
If the thermal phase is connected to the warm phase, the condition
$\Upsilon\dot{\phi}_{2I}^2 \simeq 4H \rho_R$ is satisfied in the warm phase. 
However, as far as the radiation dominates the Universe,
 it is impossible to keep $\Upsilon\dot{\phi}_{2I}^2 \simeq 4H
\rho_R$ for a time period $\Delta t \sim H^{-1}$.
In the usual warm inflationary scenario (with monomial
potential) the condition  $\Upsilon\dot{\phi}_{2I}^2 \simeq 4H\rho_R$
is equivalent to the slow-roll condition in the warm phase 
$\eta/r \sim \rho_R/V_{\phi_{2I}}<1$, 
although the equivalence is obtained using the warm-phase condition
combined with the slow-roll field equation. 
Warm inflation starts with $\eta\ll r$ in the warm phase 
and ends at $\eta\sim r$, where the slow-roll condition in the warm
phase breaks. 
Unlike the conventional warm inflation scenario, which starts with the
potential-dominated Universe in the warm phase, our
scenario starts with the radiation-dominated Universe in the thermal
phase, the warm-phase condition is violated.
Natural conclusion is that the radiation-dominated Universe is 
always in the thermal phase because the radiation energy density is too
large to be sustained by the dissipation caused by the potential. 
In fact, in the radiation-dominated Universe $\Upsilon\dot{\phi}_{2I}^2
\simeq 4H\rho_R$ may appear for a short period of time ($\Delta t \ll
H^{-1}$) but it is impossible to 
sustain the radiation against red-shifting. 
In our scenario we consider significant dissipation for the
field motion, but we may disregard the warm phase when
the radiation dominates the Universe.

For the slow-roll in the thermal phase of the radiation-dominated
Universe, we have to consider slow-roll conditions that are different
from the usual warm inflation scenario.
For the thermal phase we introduce the condition $\dot{V}< 4H V$.
Considering $\dot{V}\simeq V_{\phi_{2I}}\dot{\phi}_{2I}
\simeq \Upsilon \dot{\phi}_{2I}^2$, it leads to
\begin{eqnarray}
\frac{\dot{V}}{4H V}&\simeq& 
\left\{
\begin{array}{cccc}
\frac{n^2 V}{4C_1 T^3}
\frac{\sqrt{3}M_p}{C_T^{1/2}T^2}<1
& \,\,\,\,\, (\Upsilon_1 >\Upsilon_2)\\
\frac{n^2 V}{4C_2 \phi_{2I}^3}
\frac{\sqrt{3}M_p}{C_T^{1/2}T^2}<1
 &\,\,\,\,\,  (\Upsilon_1 <\Upsilon_2).
\end{array}
\right.
\end{eqnarray}
For the scenario with $\Upsilon_1 >\Upsilon_2$, $T\sim V^{1/4}$
leads to the condition   
$T> n^2 \sqrt{3}M_p /(4C_1 C_T^{1/2})$, which means that 
$T\sim V^{1/4}$ is consistent with the slow-roll condition only when
$T> n^2 \sqrt{3}M_p /(4C_1 C_T^{1/2})$.
Namely, if $V< n^2 \sqrt{3}M_p /(4C_1
C_T^{1/2})$, $V(\phi_{2I})$ starts decreasing faster than the
radiation before it dominates the Universe.
This scenario is not suitable for our argument and it 
will not be considered in this paper, because it can
lead neither to the entropy production nor to the generation of the
curvature perturbation. 

The dynamics related to the time-dependent friction 
($\Upsilon\propto T^3$) in the radiation-dominated Universe
is similar to the usual curvaton dynamics before oscillation, 
where the friction scales as $H\propto T^2$.
In this sense, one may say that what is considered in this section is
nothing but the dynamics of the curvatons
\cite{Curvaton-dynamics} with
the enhanced friction term $\Upsilon \gg H$.

After reheating, the Universe is in the thermal phase, where
the temperature decreases as $T\propto a^{-1}$.
Using $H^2 =\rho_R/3M_p^2$, we find for $r>1$;
\begin{eqnarray}
\frac{1}{(1+r)^2}&\simeq& 
\left\{
\begin{array}{cccc}
\frac{\rho_R \phi_{2I}^4}{M_p^2 C_1^2 T^6}
& \,\,\,\,\, (\phi_{2I}<\phi_*)\\
\frac{\rho_R }{M_p^2 C_2^2 \phi_{2I}^2}
 &\,\,\,\,\, (\phi_{2I}>\phi_*).
\end{array}
\right.
\end{eqnarray}
For the strong dissipative scenario, there are conditions 
for $r>1$, which is given by
\begin{eqnarray}
\label{rlar1}
\phi_{2I} &<& \frac{\sqrt{C_1}}{C_T^{1/4}}
\sqrt{T M_p} \,\,\,\,\, (\phi_{2I}<\phi_*)\\
\phi_{2I} &>& \frac{H}{C_2} \,\,\,\,\, (\phi_{2I}>\phi_*),
\end{eqnarray}
which are trivial for the scenario.
The slow-roll condition $\epsilon_w\le \eta_w<1$ leads to
\begin{equation}
\eta_w \simeq 
\left\{
\begin{array}{cccc}
\frac{n(n-1)C_T}{C_1^2} 
\left(\frac{\phi_{2I}}{T}\right)^2\left(\frac{V}{\rho_R}\right)<1
& \,\,\,\,\, (\phi_{2I}<\phi_*)\\
\frac{n(n-1)V}{C_2^2 \phi_{2I}^4} <1
&\,\,\,\,\, (\phi_{2I}>\phi_*),
\end{array}
\right.
\end{equation}
which result in the conditions 
\begin{eqnarray}
\label{phic}
\phi_{2I} &<& \frac{C_1 T^3}{\sqrt{n(n-1)}V^{1/2}}
\sim T \times 10^4 
\left(\frac{C_1}{10^4}\right)
\left(\frac{T}{V^{1/4}}\right)^{2}\,\,\,\,\, (\phi_{2I}<\phi_*)\\
\label{conv-rad-slo}
\phi_{2I} &>& \left[n(n-1)\right]^{1/4}V(\phi_{2I})^{1/4}
\left(\frac{1}{C_2}\right)^{1/2}\sim V(\phi_{2I})^{1/4}
\left(\frac{1}{C_2}\right)^{1/2}
\,\,\,\,\, (\phi_{2I}>\phi_*).
\end{eqnarray}
Here $\phi_*$ is estimated as
\begin{equation}
\phi_* \simeq 10 \times T \left(\frac{C_1}{10^4}\right)^{1/3}
\left(\frac{1}{C_2}\right)^{1/3}.
\end{equation}
For $\phi_{2I}<\phi_*$, the condition $\dot{V}<4HV$ in the thermal phase
puts a lower bound for the temperature.
For $\phi_{2I}>\phi_*$, it leads to the condition
\begin{equation}
\phi_{2I}> \left(\frac{n^2 V}{4C_2}
\frac{\sqrt{3}M_p}{C_T^{1/2}T^2}\right)^{1/3}
\sim V^{1/4} \left(\frac{1}{C_T^{1/2}C_2}\right)^{1/3}
\left(\frac{n^2 V^{1/4}M_p}{4T^2}\right)^{1/3},
\end{equation}
which will be more stringent than the conventional slow-roll condition
given in Eq.(\ref{conv-rad-slo}).

These equations show that the slow-roll is conceivable for the field
$\phi_{2I}$ in the radiation-dominated Universe.
The warm phase never appears during this period.
As far as the slow-roll conditions are satisfied, the curvaton field
$\phi_{2I}$ rolls slowly due to the dissipation, and the radiation
decreases as $\rho_R\propto a^{-4}$.
Then the potential $V(\phi_{2I})$ may eventually start to dominate the 
energy.
We denote the temperature by $T_{in}$, when the potential starts
domination.
The slow-roll conditions at  $T=T_{in}\simeq V^{1/4}$ give
\begin{eqnarray}
\phi_{2I}&<& 10^{4} \times
T_{in}
\left(\frac{C_1}{10^4}\right)
\,\,\,\,\, (\phi_{2I}<\phi_*)\\
\phi_{2I} &>& 
V^{1/4} \left(\frac{1}{C_T^{1/2}C_2}\right)^{1/3}
\left(\frac{n^2M_p}{4V^{1/4}}\right)^{1/3}
\,\,\,\,\, (\phi_{2I}>\phi_*).
\end{eqnarray}

\subsection{After $V(\phi_{2I})$ dominates the energy}

The potential may start to dominate the energy of the Universe as the
radiation decreases with time while the field $\phi_{2I}$ rolls slowly.
Considering the scenario in which the potential can start dominating
the Universe, our question is whether the warm phase may appear
for the dissipative curvaton $\phi_{2I}$ before the end of the slow-roll. 
Domination by the potential starts in connection with the 
thermal phase of the radiation-dominated Universe, where 
$\dot{\rho}_R \sim 4H \rho_R > \Upsilon \dot{\phi}_{2I}^2$.
Then the Universe may reach the warm phase ($\dot{\rho}_R <4H\rho_R \sim
 \Upsilon \dot{\phi}_{2I}^2$) before the end of the slow-roll.

For the slow-roll in the thermal phase, the condition 
$\dot{V}< 4H V$ leads to
\begin{eqnarray}
\frac{\dot{V}}{4H V}&\simeq& 
\left\{
\begin{array}{cccc}
\frac{n^2 \sqrt{3}M_p V^{1/2}}{4C_1 T^3}<1
& \,\,\,\,\, (\phi_{2I}<\phi_*)\\
\frac{n^2 \sqrt{3}M_pV^{1/2}}{4C_2 \phi_{2I}^3}<1
 &\,\,\,\,\,  (\phi_{2I}>\phi_*).
\end{array}
\right.
\end{eqnarray}
Conventional slow-roll conditions in the thermal phase lead to
\begin{eqnarray}
\label{phidc}
\phi_{2I} &<& \phi_{c1}\equiv  T \times 10^4 
\left(\frac{C_1}{10^4}\right)
\left(\frac{T}{V(\phi_{2I})^{1/4}}\right)^{2}\,\,\,\,\, (\phi_{2I}<\phi_*)\\
\phi_{2I} &>& \phi_{c2}\equiv V(\phi_{2I})^{1/4}
\left(\frac{1}{C_2}\right)^{1/2}
\,\,\,\,\, (\phi_{2I}>\phi_*),
\end{eqnarray}
which are identical to the slow-roll conditions in the
radiation-dominated Universe, except for 
the requirement $C_T^{1/4} T< V(\phi_{2I})^{1/4}$.

We will discuss how (and when) the warm phase appears 
in connection with the thermal phase of the potential-dominated Universe.

\underline{$\phi_{2I}<\phi_*$}

The warm phase never appears if the slow-roll in the thermal phase ends
before the warm phase begins.
For $\phi_{2I}<\phi_*$, the thermal phase ends at 
\begin{equation}
T_{end} = 
\left(\frac{n^2 \sqrt{3}M_p V(\phi_{2I})^{1/2}}{4C_1}\right)^{1/3},
\end{equation}
where the potential starts decreasing faster than the radiation.
The above condition combined with $C_T^{1/4} T<V^{1/4}$
 puts a severe bound for the curvaton potential.
Moreover, the conventional slow-roll condition shows that the upper
bound, which scales as $\phi_{c1}\propto T^3$,
decreases fast in the thermal phase.
If $\phi_{c1}$ reaches $\phi_{2I}$ before the warm phase begins,
the thermal phase ends there and the warm phase never
begins.\footnote{Equivalently, 
the slow-roll parameter increases rapidly in the thermal phase 
and eventually reach $\eta_w \sim 1$.}

The process may be delayed if the warm phase appears before the end of
the thermal phase.
In fact, the source term for the radiation is given by
\begin{equation}
\Upsilon \dot{\phi}_{2I}^2\sim 
\frac{n^2 V(\phi_{2I})^2}{9C_1T^3},
\end{equation}
 which (in the thermal phase) increases as 
$\Upsilon \dot{\phi}_{2I}^2\propto T^{-3}$.
If $\Upsilon \dot{\phi}_{2I}^2$ reaches $\Upsilon \dot{\phi}_{2I}^2\sim
4H\rho_R$ in the thermal phase, warm inflation may start and the
temperature will be sustained.
However, the large potential energy required for the scenario may not
be conceivable for the curvaton scenario.
Moreover, the chaotic initial condition for the field $\phi_{2I}$ may
favor the scenario with $\phi_{2I}>\phi_*$.

\underline{$\phi_{2I}>\phi_*$}

In contrast to the scenario with $\phi_{2I}<\phi_*$, the slow-roll
conditions combined with $\dot{V}<4HV$ 
shows that for the scenario with $\Upsilon_2>\Upsilon_1$ 
the thermal phase can be connected to the warm phase without
introducing neither large potential nor the small (and fine-tuned)
vacuum expectation value(VEV) for the curvaton field.
Once the Universe goes into the warm phase, the dynamics related to the
 slow-roll is identical to the usual warm inflation scenario
with the monomial;
\begin{equation}
V(\phi_{2I})=\lambda_n \frac{\phi_{2I}^n}{M_p^{n-4}}.
\end{equation}
The field $\phi_{2I}$ can generate the cosmological perturbation once
it dominates the Universe.
The scenario for generating cosmological perturbation
is different from the usual warm inflation scenario, because the
field fluctuations that are relevant for the cosmological perturbations 
exit horizon during primordial inflation.
The mechanism is rather similar to the curvaton scenario, although the
amplification of the energy ratio is not due to the 
long-time oscillation of the curvaton field.
The generation of the curvature perturbation will be
discussed in Sec.\ref{curv-pert}.

The scenario with $\phi_{2I} > \phi_{c2}> \phi_{*}$ allows
dissipative inflation in the thermal phase that is connected to the warm 
phase.
In the thermal phase, the lower boundary of the slow-roll region is
determined by the condition $\dot{V}<4HV$, which leads to
\begin{equation}
\label{low-warm}
\phi_{2I} > \left(\frac{n^2\sqrt{3}M_pV^{1/2}}{4C_2}\right)^{1/3}.
\end{equation}
The condition does not depend explicitly on the
temperature.\footnote{Here we have in mind the rapid change of $T$ in
the thermal phase. In the warm phase there is the relation between
$\phi_{2I}$ and $T$, where both $\phi_{2I}$ and $T$ change slowly.}
The thermal phase can be connected to the warm phase
if $\Upsilon \dot{\phi}_{2I}^2$ can reach $\Upsilon
\dot{\phi}_{2I}^2\sim 4H\rho_R$ before the slow-roll condition is
violated.
The warm phase appears at (i.e., the warm-phase condition $\Upsilon
\dot{\phi}_{2I}^2 \simeq 4H\rho_R$ is satisfied at)
\begin{equation}
\label{tw-end}
T_w\simeq \left(\frac{\alpha_2 M_p V(\phi_{2I})^{3/2}}{\phi_{2I}^3}
\right)^{1/4}
\sim 10^{9}GeV 
\left(\frac{\alpha_2}{10^{-2}}\right)^{1/4}
\left(\frac{V(\phi_{2I})^{1/4}}{10^{10}GeV}\right)^{3/2}
\left(\frac{10^{13}GeV}{\phi_{2I}}\right)^{3/4},
\end{equation}
which allows for $\alpha_2\sim 10^{-2}$, $V(\phi_{2I})^{1/4}\sim
10^{10}GeV$ and $\phi_{2I}\sim 10^{13}GeV$ the number of e-foldings of
$N^{th} \sim \ln (T_{in}/T_w)\sim 2$ in the thermal phase.
Dissipative inflation in the thermal phase, which looks like thermal
inflation without symmetry restoration, is indeed possible.

When the Universe in the thermal phase is connected to the warm phase,
warm inflation starts  
with the dissipation coefficient $\Upsilon=C_2 \phi_{2I}$.
Warm inflation ends when the radiation reaches $\rho_R\sim V(\phi_{2I})$
again, where the usual slow-roll condition for the warm inflation 
scenario is broken. 
For the monomial potential, warm inflation ends at $\eta\simeq r$,
which leads to
\begin{eqnarray}
\phi_{2I}^{(end)}&\sim & M_p 
\left(\frac{\sqrt{\lambda_n}n(n-1)}{C_2}\right)^{2/(6-n)},
\end{eqnarray}
where $n=6$ leads to a trivial result as far as the field rolls with
strong dissipation.  
For $n=6$, warm inflation may end when the strong dissipation ends.
The number of e-foldings elapsed in the warm phase is
\begin{eqnarray}
N^w &\sim& \int \frac{H}{\dot{\phi}_{2I}}d\phi_{2I}\nonumber\\
&\sim&
\left\{
\begin{array}{cc}
 \frac{2C_2}{n(6-n)\sqrt{\lambda_n}}\left[
\left(\frac{\phi_{2I}^{(ini)}}{M_p}\right)^{(6-n)/2}-
\left(\frac{\phi_{2I}^{(end)}}{M_p}\right)^{(6-n)/2}\right]
&(n\ne 6)\\
\frac{C_2}{6\sqrt{\lambda_n}}
\ln\frac{\phi^{(ini)}_{2I}}{\phi^{(end)}_{2I}}
&(n=6)
\end{array}
\right.
\end{eqnarray}
where $\phi^{(ini)}_{2I}$ denotes the value of $\phi_{2I}$ when 
the warm phase starts.
The number of e-foldings is crucially controlled by the
initial condition of the field $\phi_{2I}$.
More precisely, the field rolls slowly in the thermal phase from $T=T_R$
to $T=T_w$, but the value of $\phi_{2I}$ at the end of the thermal phase
(it corresponds to $\phi^{(ini)}_{2I}$ at the beginning of the warm phase)
is not determined by the model parameters.
It depends on the initial value of $\phi_{2I}$ during the primary inflation.
The situation may be in contrast to $\phi^{(end)}$, which is determined
by the model parameters.

Finally, we will comment on the temperature at the
beginning of the warm phase.
In conventional warm inflation the temperature is already put at the
equilibrium value from the beginning.
In contrast to the usual scenario of warm inflation,
 our scenario stars with reheating after the primary inflation,
which cannot lead to the same situation.
Namely, the initial temperature (reheating temperature  $T_R$)
cannot be identical to 
the equilibrium temperature of the warm phase($T_{eq}$).
Therefore, we must  consider  the non-equilibrium evolution of the
temperature before the warm phase.
For $T_R > T_{eq}$, the evolution after reheating is very simple, as we
have discussed for the thermal phase.
However, for $T_R < T_{eq}$, it is not clear if the temperature can
increase against red-shifting.
We understand that this problem of the initial condition is one of the
most confusing aspects of the warm inflation model.
Intuitively it seems difficult to start warm inflation from the cold
Universe.
The expectation considered in Ref.\cite{Gamma-Ms0} is that 
initially the dissipation related to the quasiparticle approximation
with zero-temperature works to raise the radiation temperature 
from $T\sim 0$ to $T=T_{eq}$.
The non-equilibrium phase will be very short if this expectation is true.
In this paper we will not discuss the scenario with $T_R<T_{eq}$.

\subsection{Curvature perturbation from the dissipative field}
\label{curv-pert}

The field $\phi_{2I}$ that rolls slowly during primordial inflation will
have the cosmological fluctuation $\delta \phi_{2I}$
that exits horizon during primordial inflation. 
If the slow-roll during primordial inflation is realized in the warm
phase, $\delta \phi_{2I}$ may be enhanced by the temperature $T>H$.
Moreover, if the potential of the field $\phi_{2I}$ dominates the energy
after reheating, it can cause generation of the cosmological
perturbation.
If the amplitude of the cosmological perturbations created by the field
$\phi_{2I}$ is larger than the one from the primordial inflation,
it gives the new source of the cosmological perturbations.

Before discussing the scenario of the dissipative curvatons, it will be
better to review the original scenario of the
curvatons \cite{curvaton-paper} to show 
clearly the differences that discriminate our scenario from the 
original.\footnote{Some other variations of the curvaton scenario are
found in Ref. \cite{matsuda_curvaton, Topolo-curv, Hybrid-Curvatons}. }
One of the crucial assumptions in the original curvaton model is that
during inflation 
the Hubble expansion rate is larger than the effective mass of the
curvaton field $\varphi$; $H_{1I}\gg m_\varphi$.
Then the quantum fluctuations are created for the curvaton field forming
a condensate with large VEV, $\varphi_0\ne 0$.
The potential of the curvaton field is negligible during inflation
since it is much smaller than the inflaton potential.
After inflation (and after reheating), the curvaton can stay at large
VEV because of the large Hubble friction.
During the radiation-dominated epoch the Hubble parameter scales as
$H\propto T^2$.
The curvaton follows the usual slow-roll equation of motion until 
$H_{osc}\equiv m_\varphi$, when the curvaton starts oscillating around
the origin.
The initial amplitude of the oscillation is given by the VEV
$\varphi_0$, which is due to the slow-roll assumption for the curvaton
field.  
During oscillation the amplitude $|\varphi|$ is redshifted by the Hubble
expansion.
Since the oscillation with the quadratic potential behaves like matter
with regard to the Hubble expansion rate $H(t)$, the energy of the
oscillating curvaton may eventually start dominating over the inflaton
decay products.\footnote{In this respect the quadratic potential is
crucial for the original curvaton model. The situation is in contrast to
the dissipative curvaton model, in which $V\propto \phi_{2I}^n$ is
possible.} 
In this scenario the longevity of the curvaton oscillation
is crucial for its dominance.
For the longevity of the curvaton, it is important that the curvaton
does not evaporate due to interactions with the thermal plasma.
In the presence of the interactions with the thermal plasma,
the oscillation may lead to efficient decay.
Therefore, contrary to our scenario, the curvaton in the original
scenario must not have interactions with the inflaton decay
products. 
The requirement for the longevity excludes many MSSM flat directions
from the curvaton candidate.\footnote{Some MSSM directions can remain
long lived \cite{long-MSSM}. However, if the inflaton decay products have
the MSSM quanta, their interactions with the MSSM curvaton may ruin the 
coherence and the longevity of the MSSM curvatons.
In some cases the isocurvature perturbation puts crucial condition
for the curvaton model \cite{iso-long,AD-iso}.} 

According to the inflation paradigm, the curvature perturbation 
that is sourced by the inflaton perturbation is constant 
(i.e., inhomogeneous but time-independent) from the end
of inflation. 
In this case the primordial curvature perturbation is given by
\begin{equation}
\zeta =\frac{H}{\dot{\phi}_{1I}}\delta \phi_{1I}.
\end{equation}
In the curvaton scenario the field responsible for the curvature
perturbation can be any field different from $\phi_{1I}$.
The curvature perturbation sourced by the curvaton $\varphi$ is given as 
a function of time by
\begin{equation}
\zeta(k,t)_\varphi=N_\varphi\delta \varphi(k),
\end{equation}
where $N$ is the number of e-folds of expansion from the horizon exit at
$k=aH$ to the time when $\zeta$ is evaluated. 
Here the subscript of $N$ denotes the derivative with respect to the
field $\varphi$.
For the original (cold) curvaton scenario the spectrum is given by
\begin{equation}
{\cal P}_\zeta^{1/2}=|N_\varphi|\frac{H_k}{2\pi},
\end{equation}
where $H_k$ is the Hubble parameter at horizon exit.
The assumption of the curvaton scenario is that the curvaton oscillation
lasts long in a background of radiation.
Usually the curvaton energy density $\rho_\varphi$ is supposed to be
negligible until long after the oscillation starts.
Quantities associated with the curvaton energy density grows during
oscillation since $\rho_\varphi/\rho_R$ is proportional to $a(t)$.
This assumption is in contrast to the dissipative curvaton scenario,
in which the curvaton energy density grows due to the dissipative
slow-roll and decays fast after oscillation begins.\footnote{The
efficient decay is supposed in the heavy curvaton
scenario \cite{matsuda_curvaton}. 
However, the heavy curvaton scenario requires a sudden growth of the
curvaton mass (or a phase transition \cite{Topolo-curv,
Hybrid-Curvatons}), which raises the curvaton energy density before the
oscillation begins.}   
To see the growth of the perturbation associated with the curvaton
field, it is useful to consider $\zeta_R\simeq 0$ for the radiation from
the inflaton decay product and consider the curvature perturbation from
the curvaton $\zeta_\varphi=\frac{1}{3} 
\frac{\delta \rho_\varphi}{\rho_\varphi}\simeq \frac{2}{3}
\frac{\delta \varphi}{\varphi}$, where the last equation is evaluated
for the quadratic potential.
The curvaton is supposed to decay before the cosmological scale enters
the horizon.
The curvature perturbation at the decay is given by
\begin{equation} 
\zeta=\frac{1}{3}\frac{\delta \rho_\varphi}{\rho+P}
\equiv r_\varphi \zeta_\varphi,
\end{equation}
where $\rho$ and $P$ are for the total energy density and the pressure
of the Universe.

Next, we consider the curvature perturbation from the original warm
inflation scenario.
For the warm inflation scenario with $T>H$, the amplitude of the thermal
fluctuation of a scalar field will be larger than the quantum
fluctuation. 
The amplitude of the spectrum in the strong dissipative scenario of 
warm inflation is given by \cite{spectrum-warm}
\begin{equation}
{\cal P}_\zeta\simeq \frac{H_k}{\dot{\phi}_{1I}}
\left[
\left(\frac{\pi r}{4}\right)^{1/4}\sqrt{T_k H_k}
\right]
\simeq 
\frac{3H_k^2r}{|V_{\phi_{1I}}|}
\left(\frac{\pi r}{4}\right)^{1/4}\sqrt{T_k H_k},
\end{equation}
where $T_k$ is the temperature when the perturbation exit horizon.
Again, it is useful to consider the curvaton scenario that is based on
the idea that the field responsible for the curvature perturbation can
be any field different from $\phi_{1I}$. 
Here we consider a dissipative field and 
denote the field by ``dissipative curvaton field'' $\phi_{2I}$.
The dissipative curvaton $\phi_{2I}$ might be one of the fields in the
(multi-field) warm inflation 
model, but in this paper $\phi_{2I}$ is supposed to be a field
which has no significant effect on the dynamics of primordial 
inflation.\footnote{Evolution of the curvature perturbation in a
multi-field warm inflation model is discussed in
Ref. \cite{matsuda-warm}. Other applications and extensions of warm
inflation scenario with regard to the multi-field inflationary model
are found in Ref. \cite{matsuda-warm-apps}.}
Whatever the origin of the field $\phi_{2I}$, the curvature perturbation
caused by the curvaton is given as a function of time by
\begin{equation}
\zeta(k,t)_{2I}=N_{\phi_{2I}}\delta \phi(k)_{2I}.
\end{equation}
The assumption of the dissipative curvaton scenario is that the curvaton
slow-roll lasts long in a background of radiation.
Usually the curvaton energy density $\rho_\varphi$ is supposed to be
negligible at reheating.
The situation is the same in our model, in which the dissipative 
curvaton energy density $\rho_{\phi_{2I}}$ is supposed to be
negligible at reheating.
Quantities associated with the curvaton energy density grows during
dissipative slow-roll epoch, where $\rho_{\phi_{2I}}/\rho_R$ is 
proportional to $a^4$.
The situation related to the slow-roll is in contrast to the original
curvaton scenario, as in the original curvaton model the energy of the
curvaton cannot dominate the Universe during the slow-roll period,
unless the VEV of the curvaton is very large ($\varphi_0 >M_p$).
To see the growth of the perturbation associated with the curvaton
field, it is useful to consider that the curvature perturbation
associated with the primordial inflation is negligible ($\zeta_R\simeq
0$), where $\zeta_R$ crucially depends on the primordial inflation model
and can be adjusted without disturbing the curvaton scenario.

For the most significant case, where the curvaton is ``warm'' during
primordial inflation (i.e., when the fluctuation of the curvaton is
enhanced by the temperature $T>H$) and then the dissipative inflationary
Universe in the thermal phase is connected to warm inflation, the total
number of e-folds elapsed during the secondary inflation stage is given by
\begin{equation}
N\simeq N^{th} + N^w.
\end{equation}
Since the temperature decreases rapidly in the thermal phase, 
we may expect that the variation of $\phi_{2I}$ during the dissipative
inflation stage is negligible for the calculation.
Then, we have $T_{in}/T_w\propto V^{1/4}\phi_{2I}^{3/4}/V^{3/8}
\propto \phi_{2I}^{(6-n)/8}$ for the dissipative inflation period.
Using these equations, $N_{\phi_{2I}}$ for the fluctuation of
$\phi_{2I}^{(ini)}$  is evaluated as
\begin{eqnarray}
N_{\phi_{2I}} 
&\simeq&
\left[\frac{6-n}{8}\frac{1}{\phi_{2I}}\right]
+
\left[ \frac{C_2}{n\sqrt{\lambda_n}M_p}
\left(\frac{\phi_{2I}}{M_p}\right)^{(4-n)/2}
\right]\nonumber\\
&\simeq&\left[\frac{6-n}{8} \frac{1}{\phi_{2I}}\right]+
\left[\frac{C_2 \phi_{2I}^2}{n\sqrt{V} M_p}\right]
\end{eqnarray}
which leads to the spectrum\footnote{See Ref. \cite{IH-pt} and 
Ref. \cite{thermal-curvgen} for more discussions for generation of
the curvature perturbation caused by the inhomogeneous phase
transition. 
The idea of inhomogeneous phase transition is a generalization of 
 precedent ieas of inhomogeneous preheating \cite{IH-pr},
inhomogeneous end of 
inflation \cite{IH-end0, IH-end1, IH-end-matsuda} and modulated 
scenarios of inflation \cite{IH-end0, IH-mod1}.
}
\begin{eqnarray}
{\cal P}_{\cal \zeta}^{1/2}(k,t)_{2I}&=&
\left[\frac{6-n}{8} \frac{1}{\phi_{2I}}+ 
\frac{C_2 \phi_{2I}^2}{n\sqrt{V}M_p} \right]
\left[
\left(\frac{\pi r}{4}\right)^{1/4}\sqrt{T_k H_k}
\right].
\end{eqnarray}
Here significant interaction between the thermal
plasma and $\phi_{2I}$ is assumed when we calculate the amplitude of
the dissipative curvaton for $T_k>H_k$.
Eq.(\ref{low-warm}) suggests that the contribution from the warm phase
(the second term) is significant for the scenario. 

\subsection{Spectral index}
Before calculating the spectral index of the model, it will be useful to
show the basic equations for the rate of change of various parameters 
that are given by\cite{spectrum-warm}
\begin{eqnarray}
\frac{1}{H}\frac{d \ln H}{dt}&=&- \frac{1}{r}\epsilon\\
\frac{1}{H}\frac{d \ln T}{dt}&=&- \frac{1}{4r}
\left(2\eta-\beta-\epsilon\right)\\
\frac{1}{H}\frac{d \ln \dot{\phi}}{dt}&=&- \frac{1}{r}
\left(\eta-\beta\right)\\
\frac{1}{H}\frac{d \ln \Upsilon}{dt}&=&- \frac{1}{r}
\beta,
\end{eqnarray}
where $\beta\equiv \frac{1}{M_p^2}\frac{\Upsilon_{\phi}V_\phi}{\Gamma V}$
is the slow-roll parameter defined for the warm phase.
Variation of $T$ is important when the perturbation $\delta \phi_{2I}$ 
that exits horizon is enhanced by the thermal effect.
(i.e., when the amplitude of $\delta \phi_{2I}$ is given by
$\left[
\left(\frac{\pi r}{4}\right)^{1/4}\sqrt{T H}
\right]$.)
The enhancement occurs when the primary inflation is warm and the
interactions of the field $\phi_{2I}$ with the radiation is significant.
As in the original curvaton model, the equation for the perturbation
$\delta \phi_{2I}$ gives 
\begin{equation}
 \ddot{\delta \phi_{2I}}+3H(1+r_{2I})\dot{\delta \phi_{2I}} 
+ V_{\phi_{2I}\phi_{2I}}=0,
\end{equation}
where $r_{2I}$ is for the dissipative motion of the field $\phi_{2I}$.
The equation leads to the variation of the perturbation 
\begin{equation}
\label{curv-2I}
\delta \phi_{2I} \propto \exp\left[
{-\frac{V_{\phi_{2I}\phi_{2I}}}{3H^2(1+r_{2I})}\Delta
 N}\right],
\end{equation}
where $\Delta N$ is the number of e-foldings elapsed after horizon
exit.\footnote{Note however that for the evolution of the inflationary
curvature perturbation the time-dependence is usually considered for
$\zeta$.  Eq.(\ref{curv-2I}) is thus important for the curvaton scenario
but it is not appropriate to consider separately the evolution of
$\delta\phi_{1I}$ when one calculates the spectral index of $\zeta$.
This distinguishes the curvaton scenario from the standard inflation model.}
Considering these equations, we find the spectral index for the
curvature perturbation created in the thermal phase is given by
\begin{equation}
n_s-1 =\left[-\frac{1}{r_{1I}}\left(\frac{\epsilon_{1I}}{4}+\frac{\eta_{1I}}{2}
+\frac{\beta_{1I}}{4} \right)
+\frac{2}{r_{2I}}\eta_{2I}\right]_{k},
\end{equation}
which is obtained with the assumption that primary inflation is warm and
the interactions with $\phi_{2I}$ are significant.
We also assumed that $\phi_{2I}$ is strongly dissipative ($r_{2I}>1$)
during primordial inflation.
The subscript $k$ is for the value of the parameters at the horizon exit.
If primary inflation is cold but the curvaton is dissipative, we find
\begin{equation}
n_s-1 =\left[-\frac{2}{r_{1I}}\epsilon_{1I}
-\frac{2}{r_{2I}}\eta_{2I}\right]_k.
\end{equation}
Note that the condition for the warm phase is not identical to the
condition for the dissipative motion.
The spectral index of the curvature perturbation created in the warm
phase of dissipative inflation is different from the conventional
warm inflation scenario. 
The crucial discrimination appears because the curvature perturbation is
created by $\delta N^w \simeq H\delta t \simeq H\delta
\phi_{2I}/\dot{\phi}_{2I}$, where  
$\delta \phi_{2I}$ is created at horizon exit during primordial
inflation but $H/\dot{\phi}_{2I}$ is the value at the beginning of the
warm phase. 
For the warm phase of dissipative inflation, 
the spectral index is thus given by
\begin{equation}
\label{warm-warm}
n_s-1 =\left[-\frac{1}{r_{1I}}\left(\frac{\epsilon_{1I}}{4}+\frac{\eta_{1I}}{2}
+\frac{\beta_{1I}}{4} \right)
+\frac{2}{r_{2I}}\eta_{2I}\right]_{k}
-\frac{1}{r_{2I}}\left(2\epsilon_{2I} +2\beta_{2I}\right).
\end{equation}
where the second term comes from 
$(H/\dot{\phi}_{2I})^2$ in the curvature perturbation and it is estimated at
the beginning of the warm phase.
Since the evolution of $\delta \phi_{2I}$ is considered separately in
the above formula, we dropped $2\eta_{2I}$ in the second term.
Results are shown in Table \ref{tab:1} for comparison, where both
non-dissipative(ND) and dissipative cases are considered for
$\phi_{2I}$.
In Table 1 we considered the specific case in which $\phi_{2I}$
interacts with the background radiation sourced by $\phi_{1I}$ during
 primordial warm inflation.
The criteria for significant dissipation ($r_{2I}>1$) 
and enhancement of the amplitude $\delta \phi_{2I}$ during primordial
warm inflation may not be identical.
Therefore, we considered as a possibility (New scenario 1) that there is
a region  where $\delta \phi_{2I}$ is
enhanced by the radiation but the curvaton field is not dissipating
($r_{2I}<1$) both during and after inflation.
This scenario is not essential for the model.
\begin{table}
\begin{tabular}{|c|c|c|}
\hline
                    & Cold primary inflation  &Warm primary inflation\\
\hline
ND $\phi_{2I}$ & $\left[2\eta_{2I}-2\epsilon_{1I}\right]_k$&
$\left[2\eta_{2I}
-\frac{1}{r_{1I}}\left(\frac{\epsilon_{1I}}{4}+\frac{\eta_{1I}}{2}
+\frac{\beta_{1I}}{4} \right)\right]_{k}$\\
& {\bf Standard Curvatons}& {\bf New scenario 1}\\
\hline
Dissipative $\phi_{2I}$ &
     $\left[\frac{2\eta_{2I}}{r_{2I}}-2\epsilon_{1I}\right]_k
-\frac{1}{r_{2I}}\left(2\epsilon_{2I} +2\beta_{2I}\right).
$&
Eq.(\ref{warm-warm})\\
& {\bf New scenario 2}& {\bf New scenario 3}\\
\hline
\end{tabular}
\caption{Spectral index}
\label{tab:1}
\end{table}
\section{Conclusions and discussions}
In this paper we considered the cosmological scenario of the
dissipative curvaton field.
Strong dissipation ($r>1$) for the curvaton field may lead to interesting
possibility that the ratio of the curvaton energy density may grow
without significant oscillation period after reheating.
Moreover, the field fluctuations of the dissipative field may be
enhanced by the warm background during primordial inflation.
We investigated the possibility that these effects may alter the usual
cosmological scenarios associated with the curvaton.
The significant difference is that the dissipative curvaton can dominate
the Universe before the curvaton begins oscillation.
The dissipative slow-roll should be considered in two distinctive
``phases'' for the radiation; thermal phase where the radiation scales
as $\rho_R\propto a^{-4}$ and the warm phase where the radiation is
sustained by the dissipation.
The inflationary expansion that is expected in the thermal phase
 is similar to thermal inflation, but the
crucial difference is that symmetry restoration is not assumed for
the dissipative inflation scenario. 

Sometimes the (original) curvaton must be hidden from the inflaton decay
products in order to avoid thermal evaporation.
The situation is in contrast to the dissipative curvaton scenario, in
which the interactions with the plasma causes strong dissipation for the
curvaton and leads to the prolonged slow-roll that may be connected to
the secondary inflaton stage.

We find that the spectral index is distinctive in the dissipative
curvaton model.
There is a benefit of the dissipative curvaton scenario.
In contrast to the original curvaton model, quadratic potential is not
needed for the curvaton field, since the growth in the oscillating
period is not essential for the dissipative model.

\section{Acknowledgment}
We wish to thank K.Shima for encouragement, and our colleagues at
Tokyo University for their kind hospitality.

\begin{appendix}
\section{Dissipative motion with $\Upsilon_2$}
The most interesting aspect of the scenario may rely on the
dissipation coefficient $\Upsilon_2$, which appears due to the
zero-temperature excitation and the decay of the mass-changing heavy
field $\chi$. 
It will therefore useful to illuminate the issue of warm inflation and
the dissipative motion with regard to the coefficient $\Upsilon_2$.
In this section we try to show intuitive and physical explanations of
why the dissipative scenario is possible with $\Upsilon_2$.

The most important point that we should first mention in this appendix
will be that basically the thermalization of the decay product is {\bf not
essential} for the dissipative motion with $\Upsilon_2$.
This discriminates our scenario from the conventional warm inflation
scenario, in which thermalization is necessary because the definition
of the warm inflation scenario is given by $T>H$.
In the scenario of dissipative curvatons the Universe is simply in the
thermal phase if thermalization delays, and (in contrast to warm
inflation) it does not mean that the scenario is not conceivable.

Perhaps the most confusing aspect of the warm inflation scenario would
be that intuitively it seems difficult to realize significant radiation
($T>H$) during inflation.
On the other hand, since the mass of the intermediate field $\chi$ is
changing during 
inflation, one intuitively knows that excitation of the field is
inevitable.
However, it is not trivial to calculate the effect of the
excitation and the decay with regard to the inflaton motion.
Therefore, calculating the effect of the excitation mode and its decay
will be the primary challenge in this direction.
For instance, much work has been devoted to this problem in terms of
analytic approximations \cite{Gamma-Ms0, Gamma-Ms1, warm-analytic}
to show that the dissipative equation motion is indeed appropriate 
for the consideration of the possible dissipation during inflation.
There are works based on numerical methods \cite{numerical-based},
which show that the approximations considered in the analytic
calculations will be correct for the model.
To be specific of the quasiparticle approximation, the issue is considered in
papers in Ref. \cite{Gamma-Ms0,Gamma-Ms1}, which show that dissipative
force in the zero-temperature background appears in the dissipative
field motion with the form of Eq.(\ref{eqofmo1st}).
Therefore, the dissipative slow-roll in the thermal phase is compelling
from the past analyses with regard to the dissipative motion in the
zero-temperature and the thermal background.
Then the difficult aspect of warm inflationary model would be in the
realization of the warm phase, which depends crucially on the
thermalization process in the light field sector that includes the light
fermion $\psi$.
In fact, the important ``assumption'' in the conventional warm inflation
scenario would be that thermalization is so efficient that the radiation 
from the decay product sources continuously the radiation density to
prevent the radiation from red-shifting. 
Here, what is important in discussing thermalization in the warm phase is
the relaxation time of the radiation, which is basically {\bf independent} of
the interactions related to the dissipative motion.
Therefore, thermalization is usually ``assumed'' in the warm inflation
scenario without mentioning the specific interactions in the light-field
sector, which can be changed without modifying the interactions related
to the dissipative motion.
For more arguments of the thermalization, see
Ref.\cite{thermalization-warm}. 

Although the relaxation time of the radiation is the important issue for
realizing the warm phase, it would be modest if we assume in this paper 
that thermalization is so efficient that the dissipation can source
radiation in the warm phase.
If the thermalization delays, only the thermal phase appears for the
dissipative curvatons.
In this case the warm phase can be neglected from the discussions.
The dissipative slow-roll ends at 
\begin{equation}
\phi_{2I} \simeq 
\phi^{(c)}_{2I}\equiv 
\left(\frac{n^2\sqrt{3}M_pV^{1/2}}{4C_2}\right)^{1/3},
\end{equation}
which determines $\rho_{\phi_{2I}}$and the temperature after the
secondary reheating caused by the curvaton decay.
Since the interactions of the dissipative curvaton are significant, it
 leads to the secondary reheating temperature 
$T_{R2}\simeq \left(\rho_{\phi_{2I}}/C_T\right)^{1/4}$.
In this paper we did not assume quadratic potential for the curvatons,
but if we assume $V(\phi_{2I})\simeq m^2 \phi_{2I}^2$, we find
\begin{equation}
\left(\phi_{2I}^{(c)}\right)^2\simeq \frac{n^2 \sqrt{3}m M_p}{4C_2}
\end{equation}
and
\begin{equation}
\rho_{\phi_{2I}}\simeq \frac{n^2 \sqrt{3}m^3 M_p}{4C_2},
\end{equation}
which leads to the secondary reheating temperature 
\begin{equation}
T_{R2} \simeq \left(\frac{n^2 \sqrt{3}m^3 M_p}{4C_2 C_T}\right)^{1/4},
\end{equation}
which is obtained for the specific case of the dissipative curvaton  for
$\Upsilon_2>\Upsilon_1$ with
quadratic potential and without thermalization.

\end{appendix}

\end{document}